\documentclass[%
 reprint,
 amsmath,amssymb,
 aps,
prb,floatfix
]{revtex4-2}

\usepackage[caption=false]{subfig}
\usepackage{graphicx}
\usepackage{dcolumn}
\usepackage{bm}
\usepackage{hyperref}

\usepackage[english]{babel}

\begin{document}

\title{Synchronization and spacetime vortices in one-dimensional driven-dissipative condensates and coupled oscillator models}

\author{John P. Moroney and Paul R. Eastham}
 \affiliation{School of Physics, Trinity College Dublin, Dublin 2, Ireland}

\date{\today}

\begin{abstract}
Driven-dissipative condensates, such as those formed from polaritons, expose how the coherence of Bose-Einstein condensates evolves far from equilibrium. We consider the phase and frequency ordering in the steady-states of a one-dimensional lattice of condensates, described by a coupled oscillator model with non-odd couplings, including both time-dependent noise and a static random potential. We present numerical results for the phase and frequency distributions, and discuss them in terms of the Kardar-Parisi-Zhang equation and the physics of spacetime vortices. We find that the nucleation of spacetime vortices causes the breakdown of the single-frequency steady-state and produces a variation in the frequency with position. Such variation would provide an experimental signature of spacetime vortices. More generally, our results expose the nature of synchronization in oscillator chains with non-odd couplings, random frequencies, and noise. 
\end{abstract}

\maketitle

\section{Introduction}

In equilibrium statics and dynamics are related through the fluctuation-dissipation theorem. Non-equilibrium systems, however, break this relationship and allow new forms of ordering\ \cite{odor_universality_2004,sieberer_dynamical_2013}. A topical example is provided by driven-dissipative Bose-Einstein condensates of exciton-polaritons\ \cite{littlewood_introduction_2017,carusotto_quantum_2013,kasprzak_boseeinstein_2006} in one-dimensional lattices\ \cite{fontaine_kardarparisizhang_2022,baboux_unstable_2018}. The phase correlations in such condensates decay exponentially in space\ \cite{gladilin_spatial_2014}, as they do in equilibrium, but are stretched exponentials in time. The correlation functions found in a recent experiment\ \cite{fontaine_kardarparisizhang_2022} agree with those predicted by the Kardar-Parisi-Zhang (KPZ) equation, which is obeyed by the phase of a driven-dissipative condensate\ \cite{altman_two-dimensional_2015,he_scaling_2015,squizzato_kardar-parisi-zhang_2018,ji_temporal_2015,deligiannis_accessing_2020,ferrier_searching_2022}.

Static disorder is ubiquitous in condensed-matter systems, and often plays a decisive role in their collective behavior. It can destroy the ordered states present in the clean limit, for example in low-dimensional magnets~\cite{imry_random-field_1975}, and lead to glassy states, for example in Bose gases~\cite{fisher_boson_1989}. Although disorder could be expected to play a similarly important role for driven-dissipative condensates, it has typically been neglected. It has been considered by some of the present authors~\cite{moroney_synchronization_2021}, among others~\cite{malpuech_bose_2007,manni_polariton_2011,thunert_cavity_2016,janot_superfluid_2013}, but we neglected the time-dependent noise that is important for the coherence properties of the condensate. In this paper, we evaluate the combined effect of disorder, of strength $\sigma$, and noise, of strength $D$, in a one-dimensional driven-dissipative condensate. We find when both the noise and disorder are non-zero, $\sigma,D\neq 0$, the characteristic single-frequency steady-state is destroyed, and the condensate acquires small variations in frequency with position. This is because the combination of disorder and noise leads to a net generation of spacetime vorticity. The inhomogeneous frequency -- or equivalently chemical potential -- of this non-equilibrium condensate is a fundamental difference compared with equilibrium.

The physics of driven-dissipative condensates is one of three closely related problems, with the others being the synchronization of coupled oscillators\ \cite{pikovskij_synchronization_2003} and the growth of interfaces\ \cite{halpin-healy_kinetic_1995}. The driven-dissipative Gross-Pitaesvkii equation for the macroscopic wavefunction in a condensate implies that the phase degree-of-freedom obeys the KPZ equation\ \cite{gladilin_spatial_2014,altman_two-dimensional_2015,manneville_phase_1996,kuramoto_chemical_1984}, originally introduced to describe a growing interface\ \cite{kardar_dynamic_1986}. The stochastic nature of the gain and loss process in the condensate gives rise to noise in the Gross-Pitaesvkii equation, which becomes the standard spacetime noise term in the KPZ equation, while a random potential gives rise to so-called columnar disorder in the KPZ equation\ \cite{szendro_localization_2007,krug_directed_1993,nattermann_diffusion_1989}, i.e., an interface growth rate which is random in space but fixed in time. For a lattice of condensates, the Gross-Pitaesvkii equation can be mapped to a coupled-oscillator model with, in general, both noise and random frequencies. Unlike the better known and studied Kuramoto model, this model includes a non-odd (cosine) coupling term, which plays an important role and allows global frequency synchronization in large systems\ \cite{moroney_synchronization_2021,blasius_quasiregular_2005}. Such Kuramoto-Sakaguchi models were introduced in Ref.\ \onlinecite{sakaguchi_mutual_1988} which along with several more recent works\ \cite{moroney_synchronization_2021,gutierrez_nonequilibrium_2023,lauter_kardar-parisi-zhang_2017,lauter_pattern_2015} uses their connection to the KPZ equation.  Here, we go beyond our analysis\ \cite{moroney_synchronization_2021} of synchronization in a Kuramoto-Sakaguchi model with random frequencies to study the impact of noise in the synchronized state. We find that it generates a small but non-zero range of time-averaged frequencies, with an unusual activated dependence related to localization effects. Our results apply not just to driven-dissipative condensates, but to the many other systems described by Kuramoto-Sakaguchi models. 

The mechanism behind the breakdown of the single-frequency steady-state is the nucleation of spacetime vortices by noise. Spacetime vortices are the topological defects of a one-dimensional phase field $\theta(x,t)$, in which the phase winds by a multiple of $2\pi$ around a closed loop in spacetime. They are not described by the KPZ equation, which is for a non-compact variable and so ignores topological defects\ \cite{sieberer_topological_2018,caputo_topological_2018}. They have been considered previously\ \cite{he_space-time_2017,lauter_kardar-parisi-zhang_2017} in the absence of disorder, and shown to modify the KPZ scaling of the correlation functions at long times, and produce a vortex turbulence phase which has yet to be observed. In these cases vortices and anti-vortices are equally likely, and the resulting states have no net vorticity on large scales. In contrast, the steady-states we find have large-scale vorticity patterns, corresponding to their inhomogeneous frequency profile. Measurements of an inhomogeneous frequency profile would provide evidence of spacetime vortices and condensate physics beyond the KPZ equation. 

\section{Model}

We consider a one-dimensional lattice of polariton condensates, in which the condensate phase on the $k^{\mathrm{th}}$ lattice site is $\theta_k(t)$, with dynamics given by the Kuramoto-Sakaguchi model \begin{equation} \frac{d\theta_j}{dt}=\sum_k J_{jk}\left[\frac{1}{\alpha}\sin(\theta_k-\theta_j)-\cos(\theta_k-\theta_j)\right] + \epsilon_j + \eta_j(t).\label{eq:kurasaka}\end{equation} This model can be derived from the Gross-Pitaevskii equation describing a lattice of condensates formed in the wells of a potential\ \cite{moroney_synchronization_2021,fontaine_kardarparisizhang_2022,ohadi_synchronization_2018}, in which case $J_{jk}$ is the real-valued Josephson coupling strength between sites $j$ and $k$, $\epsilon_j$ is the energy per particle of the condensate on site $j$, and $\alpha=\Gamma/U$ is the gain saturation parameter divided by the interaction strength. Eq. (\ref{eq:kurasaka}) is valid provided the density fluctuations are fast as well as small~\cite{moroney_synchronization_2021}. Including a dissipative part to the coupling~\cite{aleiner_radiative_2012} produces a model of the same form with a redefinition of the parameters. The dissipative couplings are small for condensates trapped in the wells of a potential, for which the tight-binding wavefunctions can be taken as real, but can be large for untrapped condensates~\cite{topfer_engineering_2021,wertz_spontaneous_2010}. Such condensates also allow the realization of time-delayed couplings~\cite{topfer_time-delay_2020}, which are beyond the scope of Eq. (\ref{eq:kurasaka}).

We consider the case of nearest-neighbor couplings, which we take to be uniform and positive, $J>0$. This is appropriate for a regular lattice where the fluctuations in the coupling will be small compared with its average value. The predominant source of disorder will then be in the site-energies, $\epsilon_j$, which we suppose have standard deviation $\sigma$. The time-dependent random driving $\eta$, which arises physically from the gain and loss processes, is Gaussian white noise with strength $D$, so that $\langle \eta_i(t)\eta_j(t^\prime)\rangle=2D\delta(t-t^\prime)\delta_{ij}$. 

While the most direct application of Eq. (\ref{eq:kurasaka}) is to lattices of coupled condensates, it can also be understood as a generalization of the KPZ equation for a single extended condensate~\cite{altman_two-dimensional_2015,gladilin_spatial_2014,manneville_phase_1996,kuramoto_chemical_1984} that incorporates the compactness of the phase $\theta$~\cite{he_space-time_2017}. In this case the lattice can be viewed as a formal device that allows vortices to be treated within a phase-only theory. Conversely, if the phase differences are small, we may expand the trigonometric functions in Eq. (\ref{eq:kurasaka}) and take the continuum limit to obtain the KPZ equation, with an additional time-independent random term from the disorder,\begin{equation} \label{eq:cont}
    \frac{\partial\theta(x,t)}{\partial t}=\frac{Ja^2}{\alpha} \frac{\partial ^2\theta}{\partial x^2}+Ja^2\left(\frac{\partial\theta}{\partial x}\right)^2+\epsilon(x)+\eta(x,t).
\end{equation} Here $a$ is the lattice constant, which will be set to one in the following. Using the Cole-Hopf transform $Z=e^{\alpha\theta}$ we can rewrite this as the imaginary-time Schr\"odinger equation for a particle in a static and a dynamic random potential \begin{align}\frac{\partial Z}{\partial t}&=\frac{J}{\alpha}\frac{\partial^2 Z}{\partial x^2}+\alpha\epsilon(x)Z+\alpha\eta(x,t)Z \label{eq:imtimeschro}\\ &=-H_0 Z+\alpha\eta(x,t)Z. \nonumber\end{align}

Before discussing the general case of Eq. (\ref{eq:kurasaka}), we recall some previous results when only one type of disorder is present. We consider, here and in the remainder of this work, only the regime $\alpha\lesssim 1$, which is appropriate for polariton condensates. In the opposite limit, $\alpha\gtrsim 1$, lattice effects dominate\ \cite{moroney_synchronization_2021} and there is a first-order transition to a disordered state\ \cite{he_space-time_2017,lauter_kardar-parisi-zhang_2017}. 

Without the noise term, Eq. (\ref{eq:kurasaka}) is the Kuramoto-Sakaguchi model for a one-dimensional system of coupled self-sustained oscillators with random natural frequencies\ \cite{sakaguchi_mutual_1988}. In contrast to the Kuramoto model, the coupling is a non-odd function of the relative phases. This allows for a globally synchronized state in which all the oscillators have a single frequency\ \cite{moroney_synchronization_2021,blasius_quasiregular_2005}, even in the limit of large numbers of oscillators. The synchronized state occurs for $\sigma<\sigma_c$, at which point there is a transition to a desynchronized state. 

The nature of the synchronized states and the form of the phase boundary\ \cite{moroney_synchronization_2021,sakaguchi_mutual_1988,blasius_quasiregular_2005,gutierrez_nonequilibrium_2023} can be understood using the mapping to the imaginary-time Schr\"odinger Eq. (\ref{eq:imtimeschro}). Its solution is $Z(x,t)=\sum_n c_n e^{-E_n t} \phi_n(x)$, where the $\phi_n$ and $E_n$ are the eigenfunctions and energies of the effective Hamiltonian $H_0$. In the long-time limit $Z$ approaches the ground state of the effective random potential $-\alpha\epsilon(x)$, which is a localized state $\phi\sim e^{-|x-x_0|/\zeta}$, at some position $x_0$, with localization length\ \cite{nattermann_diffusion_1989} $\zeta\sim (J/\alpha^2\sigma)^{2/3}$. This implies that the state is synchronized in the long-time limit -- the phase  increases at the same rate at every point in space -- and that the phase profile is a triangular function of position. An example can be seen in the topmost curve of Fig.\ \ref{fig:phaseprofiles}.  

The approach to the steady-state can also be understood in this way, because at late times $Z$ will comprise a few low-energy localized states, giving rise to a phase profile comprised of a set of triangular peaks. These grow at slightly different rates, until eventually only the fastest-growing peak, corresponding to the ground state of $H_0$, remains. (We take, without loss of generality, the ground state energy to be negative.) 

We can obtain the phase boundary for synchronization~\cite{moroney_synchronization_2021} by noting that if the gradients become too large, $|\partial_x\theta|\gtrsim 1$ the compactness of the phase becomes relevant, and Eq. (\ref{eq:kurasaka}) cannot be approximated by Eq. ({\ref{eq:cont}). Since the synchronized state has $|\partial_x\theta|\sim 1/\alpha\zeta\sim (\alpha\sigma^2/J^2)^{\frac{1}{3}}$, it occurs only below a critical disorder strength, $\sigma<\sigma_c\sim J\alpha^{-1/2}$. 

In the case where there is no static disorder, $\sigma=0$, Eq. (\ref{eq:cont}), which approximates Eq. (\ref{eq:kurasaka}), becomes the standard one-dimensional KPZ equation for a growing interface, with $\theta$ playing the role of the interface position (height). The interface is rough\ \cite{halpin-healy_kinetic_1995}, with correlation function $C(x,t)=\langle (\theta(x,t)-\theta(0,0))^2\rangle\sim t^{2/3} f(|x|/|t|^{2/3})$ where the scaling function $f(y)$ is a non-zero constant at $y\rightarrow 0$, and behaves as $f(y)\sim |y|$ as $y\rightarrow \infty$. Since the width of the interface -- corresponding to the range of phases -- grows as $\Delta\theta=\sqrt{C(0,t)}\sim t^{1/3}$, the range of time-averaged frequencies decays to zero in the long-time limit: $\Delta\omega=\Delta\theta/t\sim t^{-2/3}$. The broadening of the interface by noise does not occur fast enough to give rise to different time-averaged frequencies\ \cite{lauter_kardar-parisi-zhang_2017}. This conclusion remains unchanged on considering spacetime vortices, nucleation of which is expected to lead to diffusive behavior for the phase differences at long times\ \cite{lauter_kardar-parisi-zhang_2017,he_space-time_2017}, $\Delta\theta=\sqrt{C(0,t)}\sim t^{1/2}$, so that $\Delta\omega\sim t^{-1/2}$. 

\section{Results}

\begin{figure}
    \centering
    \includegraphics{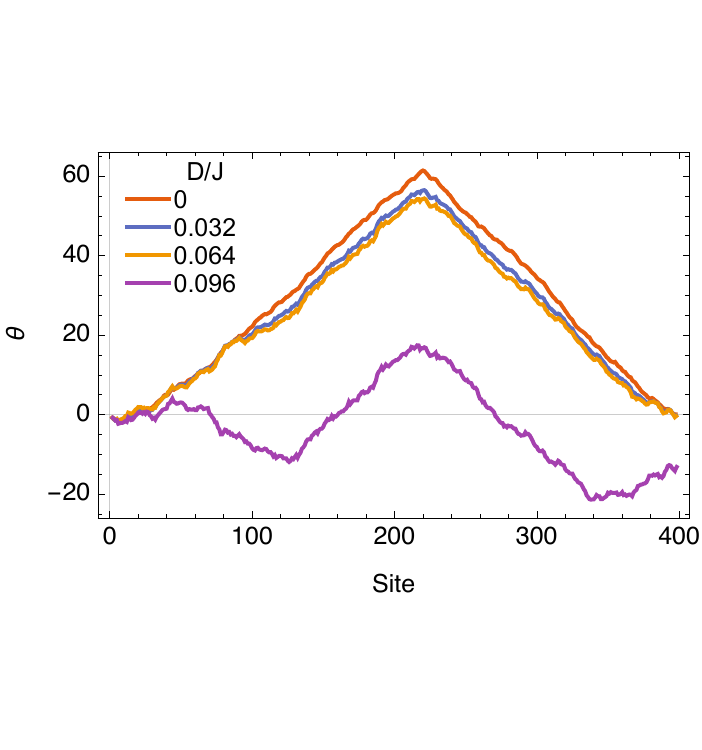}
    \caption{Phases in a chain of $400$ oscillators at the time $tJ=6000$ for various noise strengths. $\alpha=1$, $\sigma/J=0.2$, and noise strengths $D/J=0$ (red), 0.032 (blue), 0.064 (yellow), 0.128 (purple). Noise increases from top to bottom in the center of the figure. The zero of phase is chosen to be that of the first oscillator.}
    \label{fig:phaseprofiles}
\end{figure}

\subsection{Phase ordering and first-order coherence}

Fig.\ \ref{fig:phaseprofiles} shows the phases
$\theta_j(t)$ in a chain of oscillators, obtained
by integrating Eq.~(\ref{eq:kurasaka}) using a stochastic Runge-Kutta method\
\cite{rosler_rungekutta_2004}. Discontinuities in the resulting phase profiles, where neighboring phases differ by multiples of $2\pi$, have been removed to produce a smooth curve. The highest (red) curve is a typical result
obtained with disorder but without noise. The disorder strength
$\sigma=0.2J<\sigma_c$ is such that the long-time solution is
synchronized, and takes the form of a triangular phase profile as
discussed above. There are smaller variations around this overall profile, due to the residual effects of disorder\ \cite{szendro_localization_2007}. This profile is unchanged in time, apart from an overall shift. 

The remaining curves show the effects of introducing increasing
amounts of noise on the phase profiles. Qualitatively, the effect of
weak noise is to add time-dependent fluctuations about the phase
profile produced by the disorder. For the strongest noise shown, the situation is slightly more complex, with the presence of two large-scale peaks in the solution, rather than one. This corresponds to the
presence of both the ground state and the first excited state of the
effective Hamiltonian $H_0$ in the solution at this time.

We can use these observations to obtain the behavior of the first-order
coherence function of a lattice of condensates with both disorder and
noise. In polariton condensates, the coherence function $g^{(1)}(\Delta x=x-x^\prime,\Delta t=t-t^\prime)$ is determined by interfering the light emitted from one position in the lattice at one time, $(x,t)$, with that from another position at another, $(x^\prime,t^\prime)$~\cite{fontaine_kardarparisizhang_2022}. $|g^{(1)}|$ quantifies the coherence of the condensates separated by $\Delta x$ and time $\Delta t$. 

We consider the synchronized state in the regime where we may
take the continuum limit, so that there is a phase field $\theta(x,t)$
with dynamics given by Eq. (\ref{eq:cont}). Neglecting intensity fluctuations, in line with our assumptions, we have $g^{(1)}(\Delta x=x-x^\prime,\Delta t=t-t^\prime)=\langle
e^{i\theta(x,t)-i\theta(x^\prime,t^\prime)}\rangle$, where $\langle\rangle$ denotes an average.  

To study the decay of first-order coherence, we write \begin{equation}\theta(x,t)=\theta_0(x,t)+\phi(x,t),\label{eq:phasedecomp}\end{equation} where $\theta_0(x,t)$ is the steady-state solution in the absence of noise. We consider experiments done with only one realization of the random potential, which is appropriate if the static disorder arises from imperfections in the structure and only one structure is used. We also suppose that the measurement is done after the steady-state is reached. We then have $g^{(1)}=e^{i\theta_0(x,t)-i\theta_0(x^\prime,t^\prime)}\langle e^{i\phi(x,t)-i\phi(x^\prime,t^\prime)}\rangle$, where the relevant average is over noise or time but not disorder. The first factor does not fluctuate, and so does not lead to a decay of $|g^{(1)}|$, which is entirely due to the second factor. From Eq. (\ref{eq:cont}), $\phi(x,t)$ obeys \begin{align} 
    \frac{\partial\phi}{\partial t}&=\frac{J}{\alpha} \frac{\partial ^2\phi}{\partial x^2}-2J\left(\frac{\partial\theta_0}{\partial x}\right)\left(\frac{\partial\phi}{\partial x}\right)+J\left(\frac{\partial\phi}{\partial x}\right)^2+\eta(x,t) \nonumber \\ &\approx \frac{J}{\alpha} \frac{\partial ^2\phi}{\partial x^2}-2Jc\left(\frac{\partial\phi}{\partial x}\right)+J\left(\frac{\partial\phi}{\partial x}\right)^2+\eta(x,t). \label{eq:fluccont}\end{align}
In the second line we have used the fact that the steady-state solution $\theta_0$ consists of large regions where the slope is approximately constant, $\partial_x\theta_0\approx c$, and considered one such region. Eq. (\ref{eq:fluccont}) is then the standard KPZ equation with a tilted substrate\ \cite{halpin-healy_kinetic_1995}, and the second term on the right-hand side, the tilt, can be eliminated by a Galilean transformation $\tilde{x}=x-2Jct$, $\tilde{t}=t$. Thus, over each region in the solution $\theta_0$, defined by an approximately constant slope, the statistics of $\phi(\tilde{x},\tilde{t})$ and hence the decay of first-order coherence is related to that of the standard KPZ equation. 

\subsection{Desynchronization by noise}

\begin{figure}
\includegraphics{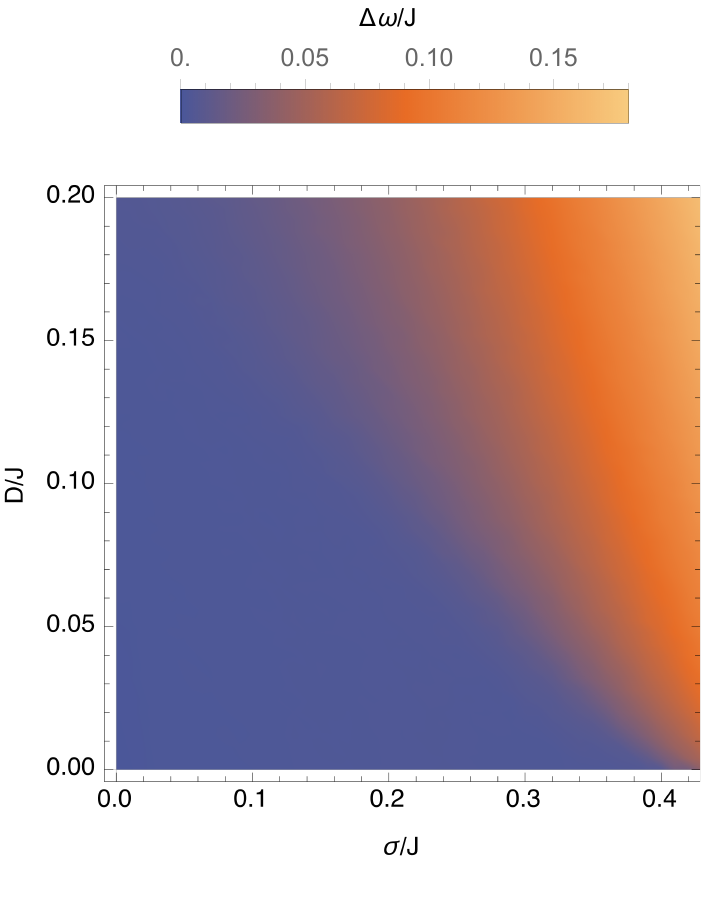}
\caption{Standard deviation of the time-averaged frequencies, $\Delta\omega$, for chains of $400$ oscillators with static disorder $\sigma$ and noise strength $D$. $\alpha=1$. The frequencies are computed over a time interval $T=1500/J$. Each point is an average of the frequency width over 32 realizations of the disorder and noise.\label{fig:freqplot}}
\end{figure}

\begin{figure}
    \centering
    \includegraphics{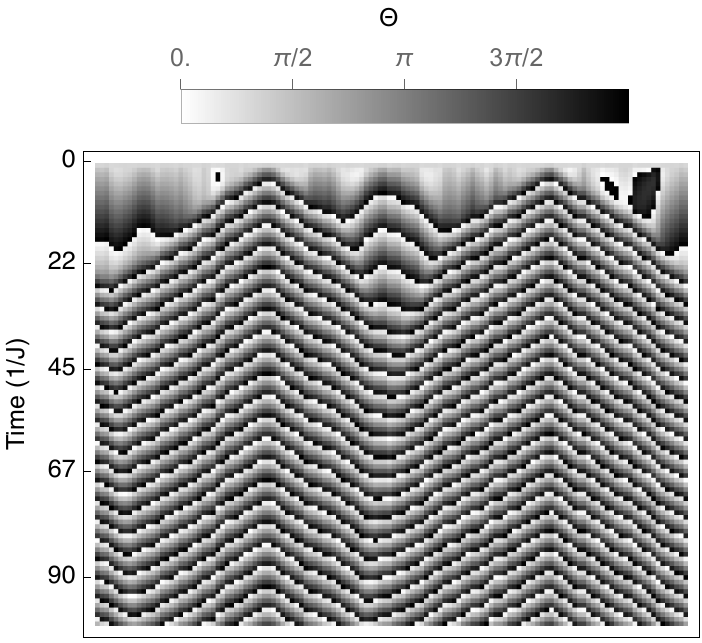}
    \includegraphics{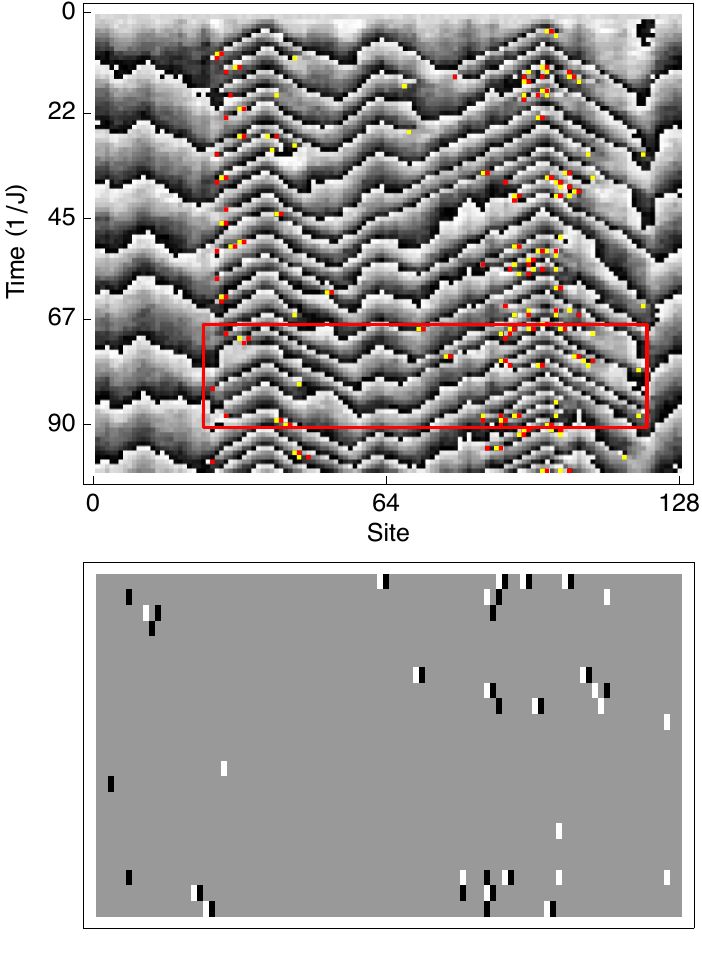}
    \caption{Phases in a chain of $N=128$ oscillators, without noise (top panel) and with noise (center panel). Position is along the horizontal axis, with time along the vertical axes, increasing from top to bottom. $\alpha=1$, $\sigma/J=0.4$, and $D=0$ (top) and $D/J=0.07$ (center). The colored points in the center panel mark spacetime vorticity +1 (red) or -1 (yellow). The bottom panel is an enlargement showing the vorticity in the region of the center panel marked with the red box, with vorticity +1 and -1 in black and white. \label{fig:examplevortices}}
\end{figure}

We now turn to consider the frequencies in the steady-state of Eq. (\ref{eq:kurasaka}). The frequency of the $k^{\mathrm{th}}$ oscillator, averaged over some long time $T$, is \begin{equation}\omega_k=[\theta_k(T+t)-\theta_k(t)]/T\label{eq:freqdef}\end{equation} This gives a frequency profile, for each realization of the disorder, whose width may be characterized by the standard deviation of the $\omega_k$, $\Delta\omega$. Fig.\ \ref{fig:freqplot} shows the results of numerical calculations of the disorder averaged width, $\langle \Delta\omega\rangle$. These results are obtained for a chain of $N=400$ oscillators, with $J=1$ and $\alpha=1$. We use a constant initial condition, and evolve to a time $t=1500$ to allow for the transients, computing the time-averaged frequencies from the phases a time $T=1500$ later.

The results along the two axes, $\sigma=0$~\cite{gladilin_spatial_2014,he_scaling_2015,he_space-time_2017,lauter_kardar-parisi-zhang_2017} and $D=0$~\cite{moroney_synchronization_2021} are expected from previous works. The state is frequency synchronized, $\Delta\omega=0$, for any $D$ when $\sigma=0$, and for $\sigma<\sigma_c\approx 0.4$ when $D=0$. However, when both noise and disorder are present we find that $\Delta\omega\neq 0$, and a range of time-averaged frequencies emerges in the solution.  

The presence of multiple time-averaged frequencies in the solution can be related to the presence of spacetime vorticity. We express the phase
change of a given site, $\theta_i(t+T)-\theta_i(t)$, as the integral
of the derivative $d\theta_i/dt$. The frequency difference between two
sites $i$ and $j$, with $i>j$, can then be expressed as a line
integral around a closed
path, \begin{align}\omega_{i}-\omega_j&=\frac{1}{T}\left(\int^{T+t}_{t}
    \frac{d\theta_i(\tau)}{d\tau} d\tau-\int^{t+T}_{t}
    \frac{d\theta_j(\tau)}{d\tau} d\tau\right) \nonumber \\ &=
  \frac{1}{T} \oint ds
  \frac{d\theta}{ds}=\frac{2\pi}{T}n_v, \label{eq:vorticityrepn} \end{align}
which counts the enclosed vorticity, $n_v$. Thus, a non-vanishing
frequency difference is equivalent to a non-vanishing density of
spacetime vorticity. The path in Eq.~(\ref{eq:vorticityrepn}) is a
rectangle starting at $(x_i,t)$ and going in the direction of
increasing time to $(x_i,t+T)$ then, in order, to $(x_j,t+T)$,
$(x_j,t)$, and back to $(x_i,t)$.  The integrals along the parts of
the path in the time direction are explicit in the first line of
Eq. (\ref{eq:vorticityrepn}), and we have chosen a gauge such that the
integrals along the space direction are zero\
\cite{he_space-time_2017}. The integrals and derivatives represent
sums and differences where they refer to a discrete coordinate.

Fig.\ \ref{fig:examplevortices} illustrates the relationship between
frequency variations and vortices. It shows the time-dependence of the
phases, with disorder alone (top), and with both disorder and noise
(center). The phase is shown over a single interval of length $2\pi$
using a grayscale, and spacetime vortices appear as dislocations in
the pattern visible in the center panel. The introduction of noise
leads to a state with a range of time-averaged frequencies, in this case a
noticeably higher frequency in the center of the chain than at the
edge. The colored dots in the center panel mark spacetime vortices with positive and
negative charges shown as different colors, and the frequency
variation along the chain can be seen to arise, as it must, from the
presence of regions with unbalanced vorticity.

\subsection{Theory of vortex nucleation}

Fig.\ \ref{fig:fwdetail} shows in more detail the computed frequency
width, as a function of the disorder strength $\sigma$, for several values of the noise $D$. We now consider the form of these
curves, in the regime $\sigma<\sigma_c$, and suggest how it can
be understood in terms of vortex nucleation.

For a first approach, we recall that the continuum description,
Eq. (\ref{eq:cont}), is based on an expansion of the trigonometric
functions in Eq.~(\ref{eq:kurasaka}), and hence becomes invalid when
the phase gradients are too large. This leads us to suggest that
vortices will be generated where the magnitude of the phase gradient
fluctuates to reach a critical value, $|\partial_x \theta|=k_c$. Furthermore, we suggest that the sign of the slope at this point corresponds to the charge of the resulting vortex. This is consistent with Fig.\ \ref{fig:examplevortices}, where we see that the positive (negative) vortices tend to occur predominantly in the regions where there is an overall positive (negative) slope of the phase profile.

As noted above, the phase profile can be decomposed as $\theta=\theta_0+\phi$, and in a region where the slope of the background is approximately constant, $\partial_x\theta_0\approx c$, the fluctuations $\phi$ obey the tilted KPZ equation, and hence the standard KPZ equation after a
Galilean transformation. Thus, the equal-time statistics of $\phi$ are identical to those of the KPZ equation, which are known to be unaffected by the nonlinear term and hence
Gaussian\ \cite{halpin-healy_kinetic_1995}. More specifically, the steady-state distribution of $\phi$ is a Gaussian with zero mean,
$P[\phi]\propto \exp\left[-\frac{J}{2D}\int dx (\partial_x
  \phi)^2\right]$. Since the slopes $\partial_x\theta=\partial_x\theta_0+\partial_x\phi\approx c+\partial_x\phi$, their distribution is this same Gaussian, shifted by $c$. The probability of a fluctuation causing the magnitude of the slope to exceed the critical value $k_c$ is then \begin{equation}P(|\partial_x\theta|>k_c)\propto \mathrm{erfc}\left(\frac{k_c\mp c}{\sqrt{2D/J}}\right)\sim e^{-(k_c\mp c)^2 J/D}, \end{equation} where the minus and plus signs in the arguments are for the cases  $\partial_x\theta>k_c$ and $\partial_x\theta <-k_c$, respectively. In a region with positive (negative) background slope, the first (second) of these will be exponentially more likely than the other in the regime of weak noise, and positive (negatively) charged vortices will predominate. This leads us to expect that a region of average slope $c$ will have a frequency width proportional to the vortex generation rate \begin{equation} G \sim e^{(2k_c-|c|)J|c|/D}.\label{eq:freqscaling}\end{equation} Since the average slope of the background scales as $c\sim1/\alpha\zeta\sim  (\alpha\sigma^2/J^2)^{1/3}$, the exponent is a sum of terms proportional to $\sigma^{2/3}$ and $\sigma^{4/3}$. The curves in the top panel of Fig.\ \ref{fig:fwdetail} are fits to a dependence of this form, which can be seen to give a good account of the data.

We have, in addition, developed a heuristic argument for the form of the noise-induced frequency width based on known results for two coupled oscillators~\cite{stratonovich_topics_1967}. For a two-site chain Eq. (\ref{eq:kurasaka}) gives \begin{equation}\frac{d}{dt}(\theta_2-\theta_1)=-\frac{2J}{\alpha}\sin(\theta_2-\theta_1)+(\epsilon_2-\epsilon_1)+(\eta_2-\eta_1).\label{eq:twooscmodel}\end{equation} This is identical to the case of Kuramoto oscillators, since the cosine term cancels. Eq. (\ref{eq:twooscmodel}) has a synchronized steady-state for detunings $\delta=|\epsilon_2-\epsilon_1|<2J/\alpha$ in the absence of noise.  The noise term nucleates phase slips in this state and introduces a frequency difference, which can be computed from the solution of the Fokker-Planck equation\ \cite{stratonovich_topics_1967}. In the present notation it is,
\begin{equation}
\delta\omega(\delta)=\delta \frac{\sinh(\pi\mu)}{\pi\mu}|I_{i\mu}(J/\alpha D)|^{-2},\label{eq:twoosc}
\end{equation} where $\mu=\delta/2D$, and $I_{i\mu}$ is a Bessel function of imaginary order. 

\begin{figure}
    \centering
    \includegraphics{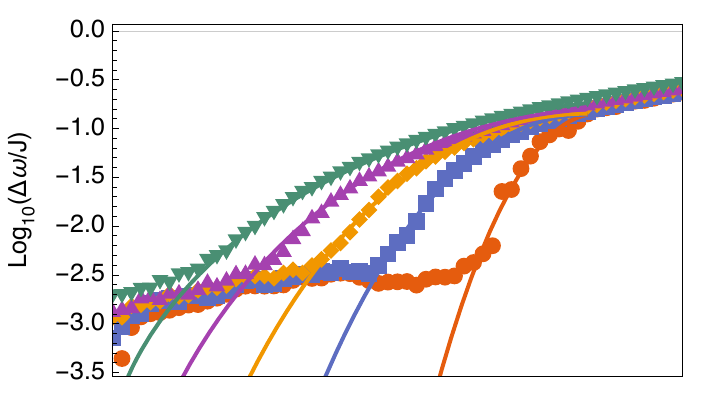}
    \includegraphics{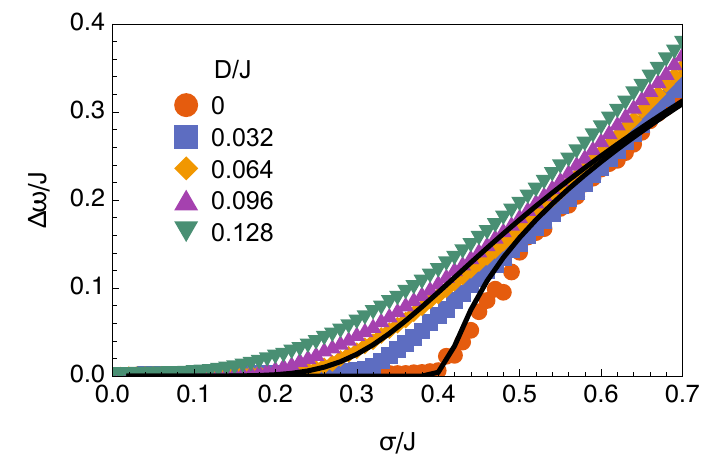}
    \caption{Dependence of the frequency width $\Delta\omega$ on disorder strength $\sigma$ for a chain of $N=400$ oscillators with $\alpha=1$ (colored points). The same data is shown on a logarithmic (top) and linear (bottom) scale. Colors indicate the noise strengths $D/J$ of 0.128 (green), 0.096 (purple), 0.064 (yellow), 0.032 (blue) and 0 (red). The curves are fits to Eq. (\ref{eq:freqscaling}) (top panel) and Eq. (\ref{eq:twooscfw}) (bottom panel), as described in the text.  \label{fig:fwdetail}}
\end{figure}

To apply this result to the non-Kuramoto chain, we recall the approximately triangular form of the background solution $\theta_0$, corresponding to the localized ground-state of the effective Hamiltonian $H_0$. This state would also be obtained for a delta-function potential of strength related to the localization length, $\epsilon(x)=-(2J/\alpha^2\zeta)\delta(x-x_0)$. More generally, a solution $\theta_0$ of saw-tooth form, arising from several low-energy states of the random potential, would be obtained from a set of such $\delta$-function potentials. This suggests associating the frequency difference $\delta$ in the coupled oscillator model with the strength of these potentials, $\delta\sim (2J/\alpha^2\zeta)\sim \alpha^{-2/3}J^{1/3}\sigma^{2/3}$. We therefore propose that the frequency width in the chain is of the form \begin{equation}\Delta\omega=C_1 \delta\omega(C_2 \alpha^{-2/3}J^{1/3}\sigma^{2/3}),\label{eq:twooscfw}\end{equation} with $\delta\omega(\delta)$ given by Eq. (\ref{eq:twoosc}). The fitting parameter $C_1$ is introduced to account for the number of sites in the chain where vortex nucleation occurs. The factor $C_2$ accounts for the details of the relationship between the localization length and the other parameters.

The lower panel of Fig.\ \ref{fig:fwdetail} shows a comparison between Eq. (\ref{eq:twooscfw}) and our simulation results. We have chosen the parameters $C_1$ and $C_2$ such that this form is close to the data in the absence of noise. We have then used these values to plot the result for $D=0.096J$. This can be seen to produce a curve close to the data (purple triangles) for that noise strength. Both curves deviate from the data in the region well above the transition, which is expected as we have neglected vortex-vortex interactions and changes in the number of sites where vortex nucleation occurs.

\section{Conclusions}

In summary, we have studied the combined effects of noise and disorder in a one-dimensional chain of driven-dissipative condensates, described by a Kuramoto-Sakaguchi oscillator model. The phase profiles, in the regimes of weak disorder and noise, consist of triangular forms produced by the disorder, with additional time-dependent fluctuations due to the noise. When spacetime vortices can be neglected these time-dependent fluctuations will be described by a tilted KPZ equation. The steady-state contains a single frequency, and the first-order coherence functions are related by a Galilean transformation to those obtained in the absence of disorder. More dramatic effects appear when spacetime vortices are considered, which lead to the breakdown of the single-frequency steady-state and the appearance of small variations in the frequency along the chain. This is due to the creation of spacetime vortices by the noise, which is biased by the currents that are induced by the disorder potential. The resulting frequency width has an unusual form, an exponential involving fractional powers of the disorder strength, reflecting the localization length of a quantum particle. 

One implication of our work is that measurements of an inhomogeneous frequency profile would provide a signature of spacetime vortices. While we have focussed on the case where such frequency variations appear due to the presence of disorder, we would expect similar effects in other potentials, since these too will induce currents in the driven-dissipative condensate that will lead to unbalanced vorticity  generation in different regions of the sample. A straightforward example would be a lattice with a single site at a different frequency, corresponding to a $\delta$-function potential in the continuum model. The mechanism we propose would also be expected to occur if a supercurrent is generated, in the absence of a potential, by imposing a phase difference between the ends of the lattice\ \cite{janot_superfluid_2013}. In this case the drop in frequency along the chain corresponds to dissipation in the supercurrent due to spacetime vortices.

\begin{acknowledgments}
We acknowledge funding from the Irish Research Council (GOIPG/2019/2824) and Science Foundation Ireland (21/FFP-P/10142). 
\end{acknowledgments} 

\bibliography{references}

\begin{thebibliography}{44}%
\makeatletter
\providecommand \@ifxundefined [1]{%
 \@ifx{#1\undefined}
}%
\providecommand \@ifnum [1]{%
 \ifnum #1\expandafter \@firstoftwo
 \else \expandafter \@secondoftwo
 \fi
}%
\providecommand \@ifx [1]{%
 \ifx #1\expandafter \@firstoftwo
 \else \expandafter \@secondoftwo
 \fi
}%
\providecommand \natexlab [1]{#1}%
\providecommand \enquote  [1]{``#1''}%
\providecommand \bibnamefont  [1]{#1}%
\providecommand \bibfnamefont [1]{#1}%
\providecommand \citenamefont [1]{#1}%
\providecommand \href@noop [0]{\@secondoftwo}%
\providecommand \href [0]{\begingroup \@sanitize@url \@href}%
\providecommand \@href[1]{\@@startlink{#1}\@@href}%
\providecommand \@@href[1]{\endgroup#1\@@endlink}%
\providecommand \@sanitize@url [0]{\catcode `\\12\catcode `\$12\catcode
  `\&12\catcode `\#12\catcode `\^12\catcode `\_12\catcode `\%12\relax}%
\providecommand \@@startlink[1]{}%
\providecommand \@@endlink[0]{}%
\providecommand \url  [0]{\begingroup\@sanitize@url \@url }%
\providecommand \@url [1]{\endgroup\@href {#1}{\urlprefix }}%
\providecommand \urlprefix  [0]{URL }%
\providecommand \Eprint [0]{\href }%
\providecommand \doibase [0]{https://doi.org/}%
\providecommand \selectlanguage [0]{\@gobble}%
\providecommand \bibinfo  [0]{\@secondoftwo}%
\providecommand \bibfield  [0]{\@secondoftwo}%
\providecommand \translation [1]{[#1]}%
\providecommand \BibitemOpen [0]{}%
\providecommand \bibitemStop [0]{}%
\providecommand \bibitemNoStop [0]{.\EOS\space}%
\providecommand \EOS [0]{\spacefactor3000\relax}%
\providecommand \BibitemShut  [1]{\csname bibitem#1\endcsname}%
\let\auto@bib@innerbib\@empty
\bibitem [{\citenamefont {Odor}(2004)}]{odor_universality_2004}%
  \BibitemOpen
  \bibfield  {author} {\bibinfo {author} {\bibfnamefont {G.}~\bibnamefont
  {Odor}},\ }\bibfield  {title} {\bibinfo {title} {Universality classes in
  nonequilibrium lattice systems},\ }\href
  {https://doi.org/10.1103/RevModPhys.76.663} {\bibfield  {journal} {\bibinfo
  {journal} {Rev. Mod. Phys.}\ }\textbf {\bibinfo {volume} {76}},\ \bibinfo
  {pages} {663} (\bibinfo {year} {2004})}\BibitemShut {NoStop}%
\bibitem [{\citenamefont {Sieberer}\ \emph {et~al.}(2013)\citenamefont
  {Sieberer}, \citenamefont {Huber}, \citenamefont {Altman},\ and\
  \citenamefont {Diehl}}]{sieberer_dynamical_2013}%
  \BibitemOpen
  \bibfield  {author} {\bibinfo {author} {\bibfnamefont {L.~M.}\ \bibnamefont
  {Sieberer}}, \bibinfo {author} {\bibfnamefont {S.~D.}\ \bibnamefont {Huber}},
  \bibinfo {author} {\bibfnamefont {E.}~\bibnamefont {Altman}},\ and\ \bibinfo
  {author} {\bibfnamefont {S.}~\bibnamefont {Diehl}},\ }\bibfield  {title}
  {\bibinfo {title} {Dynamical {{Critical Phenomena}} in {{Driven-Dissipative
  Systems}}},\ }\href {https://doi.org/10.1103/PhysRevLett.110.195301}
  {\bibfield  {journal} {\bibinfo  {journal} {Phys. Rev. Lett.}\ }\textbf
  {\bibinfo {volume} {110}},\ \bibinfo {pages} {195301} (\bibinfo {year}
  {2013})}\BibitemShut {NoStop}%
\bibitem [{\citenamefont {Littlewood}\ and\ \citenamefont
  {Edelman}(2017)}]{littlewood_introduction_2017}%
  \BibitemOpen
  \bibfield  {author} {\bibinfo {author} {\bibfnamefont {P.~B.}\ \bibnamefont
  {Littlewood}}\ and\ \bibinfo {author} {\bibfnamefont {A.}~\bibnamefont
  {Edelman}},\ }\bibfield  {title} {\bibinfo {title} {Introduction to
  {{Polariton Condensation}}},\ }in\ \href
  {https://doi.org/10.1017/9781316084366.006} {\emph {\bibinfo {booktitle}
  {Universal {{Themes}} of {{Bose-Einstein Condensation}}}}},\ \bibinfo
  {editor} {edited by\ \bibinfo {editor} {\bibfnamefont {N.~P.}\ \bibnamefont
  {Proukakis}}, \bibinfo {editor} {\bibfnamefont {D.~W.}\ \bibnamefont
  {Snoke}},\ and\ \bibinfo {editor} {\bibfnamefont {P.~B.}\ \bibnamefont
  {Littlewood}}}\ (\bibinfo  {publisher} {{Cambridge University Press}},\
  \bibinfo {year} {2017})\ pp.\ \bibinfo {pages} {57--74}\BibitemShut {NoStop}%
\bibitem [{\citenamefont {Carusotto}\ and\ \citenamefont
  {Ciuti}(2013)}]{carusotto_quantum_2013}%
  \BibitemOpen
  \bibfield  {author} {\bibinfo {author} {\bibfnamefont {I.}~\bibnamefont
  {Carusotto}}\ and\ \bibinfo {author} {\bibfnamefont {C.}~\bibnamefont
  {Ciuti}},\ }\bibfield  {title} {\bibinfo {title} {Quantum fluids of light},\
  }\href {https://doi.org/10.1103/RevModPhys.85.299} {\bibfield  {journal}
  {\bibinfo  {journal} {Rev. Mod. Phys.}\ }\textbf {\bibinfo {volume} {85}},\
  \bibinfo {pages} {299} (\bibinfo {year} {2013})}\BibitemShut {NoStop}%
\bibitem [{\citenamefont {Kasprzak}\ \emph {et~al.}(2006)\citenamefont
  {Kasprzak}, \citenamefont {Richard}, \citenamefont {Kundermann},
  \citenamefont {Baas}, \citenamefont {Jeambrun}, \citenamefont {Keeling},
  \citenamefont {Marchetti}, \citenamefont {Szyma{\'n}ska}, \citenamefont
  {Andr{\'e}}, \citenamefont {Staehli}, \citenamefont {Savona}, \citenamefont
  {Littlewood}, \citenamefont {Deveaud},\ and\ \citenamefont
  {Dang}}]{kasprzak_boseeinstein_2006}%
  \BibitemOpen
  \bibfield  {author} {\bibinfo {author} {\bibfnamefont {J.}~\bibnamefont
  {Kasprzak}}, \bibinfo {author} {\bibfnamefont {M.}~\bibnamefont {Richard}},
  \bibinfo {author} {\bibfnamefont {S.}~\bibnamefont {Kundermann}}, \bibinfo
  {author} {\bibfnamefont {A.}~\bibnamefont {Baas}}, \bibinfo {author}
  {\bibfnamefont {P.}~\bibnamefont {Jeambrun}}, \bibinfo {author}
  {\bibfnamefont {J.~M.~J.}\ \bibnamefont {Keeling}}, \bibinfo {author}
  {\bibfnamefont {F.~M.}\ \bibnamefont {Marchetti}}, \bibinfo {author}
  {\bibfnamefont {M.~H.}\ \bibnamefont {Szyma{\'n}ska}}, \bibinfo {author}
  {\bibfnamefont {R.}~\bibnamefont {Andr{\'e}}}, \bibinfo {author}
  {\bibfnamefont {J.~L.}\ \bibnamefont {Staehli}}, \bibinfo {author}
  {\bibfnamefont {V.}~\bibnamefont {Savona}}, \bibinfo {author} {\bibfnamefont
  {P.~B.}\ \bibnamefont {Littlewood}}, \bibinfo {author} {\bibfnamefont
  {B.}~\bibnamefont {Deveaud}},\ and\ \bibinfo {author} {\bibfnamefont {L.~S.}\
  \bibnamefont {Dang}},\ }\bibfield  {title} {\bibinfo {title}
  {Bose\textendash{{Einstein}} condensation of exciton polaritons},\ }\href
  {https://doi.org/10.1038/nature05131} {\bibfield  {journal} {\bibinfo
  {journal} {Nature}\ }\textbf {\bibinfo {volume} {443}},\ \bibinfo {pages}
  {409} (\bibinfo {year} {2006})}\BibitemShut {NoStop}%
\bibitem [{\citenamefont {Fontaine}\ \emph {et~al.}(2022)\citenamefont
  {Fontaine}, \citenamefont {Squizzato}, \citenamefont {Baboux}, \citenamefont
  {Amelio}, \citenamefont {Lema{\^i}tre}, \citenamefont {Morassi},
  \citenamefont {Sagnes}, \citenamefont {Le~Gratiet}, \citenamefont {Harouri},
  \citenamefont {Wouters}, \citenamefont {Carusotto}, \citenamefont {Amo},
  \citenamefont {Richard}, \citenamefont {Minguzzi}, \citenamefont {Canet},
  \citenamefont {Ravets},\ and\ \citenamefont
  {Bloch}}]{fontaine_kardarparisizhang_2022}%
  \BibitemOpen
  \bibfield  {author} {\bibinfo {author} {\bibfnamefont {Q.}~\bibnamefont
  {Fontaine}}, \bibinfo {author} {\bibfnamefont {D.}~\bibnamefont {Squizzato}},
  \bibinfo {author} {\bibfnamefont {F.}~\bibnamefont {Baboux}}, \bibinfo
  {author} {\bibfnamefont {I.}~\bibnamefont {Amelio}}, \bibinfo {author}
  {\bibfnamefont {A.}~\bibnamefont {Lema{\^i}tre}}, \bibinfo {author}
  {\bibfnamefont {M.}~\bibnamefont {Morassi}}, \bibinfo {author} {\bibfnamefont
  {I.}~\bibnamefont {Sagnes}}, \bibinfo {author} {\bibfnamefont
  {L.}~\bibnamefont {Le~Gratiet}}, \bibinfo {author} {\bibfnamefont
  {A.}~\bibnamefont {Harouri}}, \bibinfo {author} {\bibfnamefont
  {M.}~\bibnamefont {Wouters}}, \bibinfo {author} {\bibfnamefont
  {I.}~\bibnamefont {Carusotto}}, \bibinfo {author} {\bibfnamefont
  {A.}~\bibnamefont {Amo}}, \bibinfo {author} {\bibfnamefont {M.}~\bibnamefont
  {Richard}}, \bibinfo {author} {\bibfnamefont {A.}~\bibnamefont {Minguzzi}},
  \bibinfo {author} {\bibfnamefont {L.}~\bibnamefont {Canet}}, \bibinfo
  {author} {\bibfnamefont {S.}~\bibnamefont {Ravets}},\ and\ \bibinfo {author}
  {\bibfnamefont {J.}~\bibnamefont {Bloch}},\ }\bibfield  {title} {\bibinfo
  {title} {Kardar\textendash{{Parisi}}\textendash{{Zhang}} universality in a
  one-dimensional polariton condensate},\ }\href
  {https://doi.org/10.1038/s41586-022-05001-8} {\bibfield  {journal} {\bibinfo
  {journal} {Nature}\ }\textbf {\bibinfo {volume} {608}},\ \bibinfo {pages}
  {687} (\bibinfo {year} {2022})}\BibitemShut {NoStop}%
\bibitem [{\citenamefont {Baboux}\ \emph {et~al.}(2018)\citenamefont {Baboux},
  \citenamefont {Bernardis}, \citenamefont {Goblot}, \citenamefont {Gladilin},
  \citenamefont {Gomez}, \citenamefont {Galopin}, \citenamefont {Gratiet},
  \citenamefont {Lema{\^i}tre}, \citenamefont {Sagnes}, \citenamefont
  {Carusotto}, \citenamefont {Wouters}, \citenamefont {Amo},\ and\
  \citenamefont {Bloch}}]{baboux_unstable_2018}%
  \BibitemOpen
  \bibfield  {author} {\bibinfo {author} {\bibfnamefont {F.}~\bibnamefont
  {Baboux}}, \bibinfo {author} {\bibfnamefont {D.~D.}\ \bibnamefont
  {Bernardis}}, \bibinfo {author} {\bibfnamefont {V.}~\bibnamefont {Goblot}},
  \bibinfo {author} {\bibfnamefont {V.~N.}\ \bibnamefont {Gladilin}}, \bibinfo
  {author} {\bibfnamefont {C.}~\bibnamefont {Gomez}}, \bibinfo {author}
  {\bibfnamefont {E.}~\bibnamefont {Galopin}}, \bibinfo {author} {\bibfnamefont
  {L.~L.}\ \bibnamefont {Gratiet}}, \bibinfo {author} {\bibfnamefont
  {A.}~\bibnamefont {Lema{\^i}tre}}, \bibinfo {author} {\bibfnamefont
  {I.}~\bibnamefont {Sagnes}}, \bibinfo {author} {\bibfnamefont
  {I.}~\bibnamefont {Carusotto}}, \bibinfo {author} {\bibfnamefont
  {M.}~\bibnamefont {Wouters}}, \bibinfo {author} {\bibfnamefont
  {A.}~\bibnamefont {Amo}},\ and\ \bibinfo {author} {\bibfnamefont
  {J.}~\bibnamefont {Bloch}},\ }\bibfield  {title} {\bibinfo {title} {Unstable
  and stable regimes of polariton condensation},\ }\href
  {https://doi.org/10.1364/OPTICA.5.001163} {\bibfield  {journal} {\bibinfo
  {journal} {Optica}\ }\textbf {\bibinfo {volume} {5}},\ \bibinfo {pages}
  {1163} (\bibinfo {year} {2018})}\BibitemShut {NoStop}%
\bibitem [{\citenamefont {Gladilin}\ \emph {et~al.}(2014)\citenamefont
  {Gladilin}, \citenamefont {Ji},\ and\ \citenamefont
  {Wouters}}]{gladilin_spatial_2014}%
  \BibitemOpen
  \bibfield  {author} {\bibinfo {author} {\bibfnamefont {V.~N.}\ \bibnamefont
  {Gladilin}}, \bibinfo {author} {\bibfnamefont {K.}~\bibnamefont {Ji}},\ and\
  \bibinfo {author} {\bibfnamefont {M.}~\bibnamefont {Wouters}},\ }\bibfield
  {title} {\bibinfo {title} {Spatial coherence of weakly interacting
  one-dimensional nonequilibrium bosonic quantum fluids},\ }\href
  {https://doi.org/10.1103/PhysRevA.90.023615} {\bibfield  {journal} {\bibinfo
  {journal} {Phys. Rev. A}\ }\textbf {\bibinfo {volume} {90}},\ \bibinfo
  {pages} {023615} (\bibinfo {year} {2014})}\BibitemShut {NoStop}%
\bibitem [{\citenamefont {Altman}\ \emph {et~al.}(2015)\citenamefont {Altman},
  \citenamefont {Sieberer}, \citenamefont {Chen}, \citenamefont {Diehl},\ and\
  \citenamefont {Toner}}]{altman_two-dimensional_2015}%
  \BibitemOpen
  \bibfield  {author} {\bibinfo {author} {\bibfnamefont {E.}~\bibnamefont
  {Altman}}, \bibinfo {author} {\bibfnamefont {L.~M.}\ \bibnamefont
  {Sieberer}}, \bibinfo {author} {\bibfnamefont {L.}~\bibnamefont {Chen}},
  \bibinfo {author} {\bibfnamefont {S.}~\bibnamefont {Diehl}},\ and\ \bibinfo
  {author} {\bibfnamefont {J.}~\bibnamefont {Toner}},\ }\bibfield  {title}
  {\bibinfo {title} {Two-dimensional superfluidity of exciton polaritons
  requires strong anisotropy},\ }\href
  {https://doi.org/10.1103/PhysRevX.5.011017} {\bibfield  {journal} {\bibinfo
  {journal} {Phys. Rev. X}\ }\textbf {\bibinfo {volume} {5}},\ \bibinfo {pages}
  {011017} (\bibinfo {year} {2015})}\BibitemShut {NoStop}%
\bibitem [{\citenamefont {He}\ \emph {et~al.}(2015)\citenamefont {He},
  \citenamefont {Sieberer}, \citenamefont {Altman},\ and\ \citenamefont
  {Diehl}}]{he_scaling_2015}%
  \BibitemOpen
  \bibfield  {author} {\bibinfo {author} {\bibfnamefont {L.}~\bibnamefont
  {He}}, \bibinfo {author} {\bibfnamefont {L.~M.}\ \bibnamefont {Sieberer}},
  \bibinfo {author} {\bibfnamefont {E.}~\bibnamefont {Altman}},\ and\ \bibinfo
  {author} {\bibfnamefont {S.}~\bibnamefont {Diehl}},\ }\bibfield  {title}
  {\bibinfo {title} {Scaling properties of one-dimensional driven-dissipative
  condensates},\ }\href {https://doi.org/10.1103/PhysRevB.92.155307} {\bibfield
   {journal} {\bibinfo  {journal} {Phys. Rev. B}\ }\textbf {\bibinfo {volume}
  {92}},\ \bibinfo {pages} {155307} (\bibinfo {year} {2015})}\BibitemShut
  {NoStop}%
\bibitem [{\citenamefont {Squizzato}\ \emph {et~al.}(2018)\citenamefont
  {Squizzato}, \citenamefont {Canet},\ and\ \citenamefont
  {Minguzzi}}]{squizzato_kardar-parisi-zhang_2018}%
  \BibitemOpen
  \bibfield  {author} {\bibinfo {author} {\bibfnamefont {D.}~\bibnamefont
  {Squizzato}}, \bibinfo {author} {\bibfnamefont {L.}~\bibnamefont {Canet}},\
  and\ \bibinfo {author} {\bibfnamefont {A.}~\bibnamefont {Minguzzi}},\
  }\bibfield  {title} {\bibinfo {title} {Kardar-{{Parisi-Zhang}} universality
  in the phase distributions of one-dimensional exciton-polaritons},\ }\href
  {https://doi.org/10.1103/PhysRevB.97.195453} {\bibfield  {journal} {\bibinfo
  {journal} {Phys. Rev. B}\ }\textbf {\bibinfo {volume} {97}},\ \bibinfo
  {pages} {195453} (\bibinfo {year} {2018})}\BibitemShut {NoStop}%
\bibitem [{\citenamefont {Ji}\ \emph {et~al.}(2015)\citenamefont {Ji},
  \citenamefont {Gladilin},\ and\ \citenamefont {Wouters}}]{ji_temporal_2015}%
  \BibitemOpen
  \bibfield  {author} {\bibinfo {author} {\bibfnamefont {K.}~\bibnamefont
  {Ji}}, \bibinfo {author} {\bibfnamefont {V.~N.}\ \bibnamefont {Gladilin}},\
  and\ \bibinfo {author} {\bibfnamefont {M.}~\bibnamefont {Wouters}},\
  }\bibfield  {title} {\bibinfo {title} {Temporal coherence of one-dimensional
  nonequilibrium quantum fluids},\ }\href
  {https://doi.org/10.1103/PhysRevB.91.045301} {\bibfield  {journal} {\bibinfo
  {journal} {Phys. Rev. B}\ }\textbf {\bibinfo {volume} {91}},\ \bibinfo
  {pages} {045301} (\bibinfo {year} {2015})}\BibitemShut {NoStop}%
\bibitem [{\citenamefont {Deligiannis}\ \emph {et~al.}(2020)\citenamefont
  {Deligiannis}, \citenamefont {Squizzato}, \citenamefont {Minguzzi},\ and\
  \citenamefont {Canet}}]{deligiannis_accessing_2020}%
  \BibitemOpen
  \bibfield  {author} {\bibinfo {author} {\bibfnamefont {K.}~\bibnamefont
  {Deligiannis}}, \bibinfo {author} {\bibfnamefont {D.}~\bibnamefont
  {Squizzato}}, \bibinfo {author} {\bibfnamefont {A.}~\bibnamefont
  {Minguzzi}},\ and\ \bibinfo {author} {\bibfnamefont {L.}~\bibnamefont
  {Canet}},\ }\bibfield  {title} {\bibinfo {title} {Accessing
  {{Kardar-Parisi-Zhang}} universality sub-classes with exciton polaritons},\
  }\href {https://doi.org/10.1209/0295-5075/132/67004} {\bibfield  {journal}
  {\bibinfo  {journal} {EPL}\ }\textbf {\bibinfo {volume} {132}},\ \bibinfo
  {pages} {67004} (\bibinfo {year} {2020})}\BibitemShut {NoStop}%
\bibitem [{\citenamefont {Ferrier}\ \emph {et~al.}(2022)\citenamefont
  {Ferrier}, \citenamefont {Zamora}, \citenamefont {Dagvadorj},\ and\
  \citenamefont {Szyma{\'n}ska}}]{ferrier_searching_2022}%
  \BibitemOpen
  \bibfield  {author} {\bibinfo {author} {\bibfnamefont {A.}~\bibnamefont
  {Ferrier}}, \bibinfo {author} {\bibfnamefont {A.}~\bibnamefont {Zamora}},
  \bibinfo {author} {\bibfnamefont {G.}~\bibnamefont {Dagvadorj}},\ and\
  \bibinfo {author} {\bibfnamefont {M.~H.}\ \bibnamefont {Szyma{\'n}ska}},\
  }\bibfield  {title} {\bibinfo {title} {Searching for the
  {{Kardar-Parisi-Zhang}} phase in microcavity polaritons},\ }\href
  {https://doi.org/10.1103/PhysRevB.105.205301} {\bibfield  {journal} {\bibinfo
   {journal} {Phys. Rev. B}\ }\textbf {\bibinfo {volume} {105}},\ \bibinfo
  {pages} {205301} (\bibinfo {year} {2022})}\BibitemShut {NoStop}%
\bibitem [{\citenamefont {Imry}\ and\ \citenamefont
  {Ma}(1975)}]{imry_random-field_1975}%
  \BibitemOpen
  \bibfield  {author} {\bibinfo {author} {\bibfnamefont {Y.}~\bibnamefont
  {Imry}}\ and\ \bibinfo {author} {\bibfnamefont {S.-K.}\ \bibnamefont {Ma}},\
  }\bibfield  {title} {\bibinfo {title} {Random-{{Field Instability}} of the
  {{Ordered State}} of {{Continuous Symmetry}}},\ }\href
  {https://doi.org/10.1103/PhysRevLett.35.1399} {\bibfield  {journal} {\bibinfo
   {journal} {Phys. Rev. Lett.}\ }\textbf {\bibinfo {volume} {35}},\ \bibinfo
  {pages} {1399} (\bibinfo {year} {1975})}\BibitemShut {NoStop}%
\bibitem [{\citenamefont {Fisher}\ \emph {et~al.}(1989)\citenamefont {Fisher},
  \citenamefont {Weichman}, \citenamefont {Grinstein},\ and\ \citenamefont
  {Fisher}}]{fisher_boson_1989}%
  \BibitemOpen
  \bibfield  {author} {\bibinfo {author} {\bibfnamefont {M.~P.~A.}\
  \bibnamefont {Fisher}}, \bibinfo {author} {\bibfnamefont {P.~B.}\
  \bibnamefont {Weichman}}, \bibinfo {author} {\bibfnamefont {G.}~\bibnamefont
  {Grinstein}},\ and\ \bibinfo {author} {\bibfnamefont {D.~S.}\ \bibnamefont
  {Fisher}},\ }\bibfield  {title} {\bibinfo {title} {Boson localization and the
  superfluid-insulator transition},\ }\href
  {https://doi.org/10.1103/PhysRevB.40.546} {\bibfield  {journal} {\bibinfo
  {journal} {Phys. Rev. B}\ }\textbf {\bibinfo {volume} {40}},\ \bibinfo
  {pages} {546} (\bibinfo {year} {1989})}\BibitemShut {NoStop}%
\bibitem [{\citenamefont {Moroney}\ and\ \citenamefont
  {Eastham}(2021)}]{moroney_synchronization_2021}%
  \BibitemOpen
  \bibfield  {author} {\bibinfo {author} {\bibfnamefont {J.~P.}\ \bibnamefont
  {Moroney}}\ and\ \bibinfo {author} {\bibfnamefont {P.~R.}\ \bibnamefont
  {Eastham}},\ }\bibfield  {title} {\bibinfo {title} {Synchronization in
  disordered oscillator lattices: {{Nonequilibrium}} phase transition for
  driven-dissipative bosons},\ }\href
  {https://doi.org/10.1103/PhysRevResearch.3.043092} {\bibfield  {journal}
  {\bibinfo  {journal} {Phys. Rev. Research}\ }\textbf {\bibinfo {volume}
  {3}},\ \bibinfo {pages} {043092} (\bibinfo {year} {2021})}\BibitemShut
  {NoStop}%
\bibitem [{\citenamefont {Malpuech}\ \emph {et~al.}(2007)\citenamefont
  {Malpuech}, \citenamefont {Solnyshkov}, \citenamefont {Ouerdane},
  \citenamefont {Glazov},\ and\ \citenamefont {Shelykh}}]{malpuech_bose_2007}%
  \BibitemOpen
  \bibfield  {author} {\bibinfo {author} {\bibfnamefont {G.}~\bibnamefont
  {Malpuech}}, \bibinfo {author} {\bibfnamefont {D.~D.}\ \bibnamefont
  {Solnyshkov}}, \bibinfo {author} {\bibfnamefont {H.}~\bibnamefont
  {Ouerdane}}, \bibinfo {author} {\bibfnamefont {M.~M.}\ \bibnamefont
  {Glazov}},\ and\ \bibinfo {author} {\bibfnamefont {I.}~\bibnamefont
  {Shelykh}},\ }\bibfield  {title} {\bibinfo {title} {Bose {{Glass}} and
  {{Superfluid Phases}} of {{Cavity Polaritons}}},\ }\href
  {https://doi.org/10.1103/PhysRevLett.98.206402} {\bibfield  {journal}
  {\bibinfo  {journal} {Phys. Rev. Lett.}\ }\textbf {\bibinfo {volume} {98}},\
  \bibinfo {pages} {206402} (\bibinfo {year} {2007})}\BibitemShut {NoStop}%
\bibitem [{\citenamefont {Manni}\ \emph {et~al.}(2011)\citenamefont {Manni},
  \citenamefont {Lagoudakis}, \citenamefont {Pietka}, \citenamefont
  {Fontanesi}, \citenamefont {Wouters}, \citenamefont {Savona}, \citenamefont
  {Andr{\'e}},\ and\ \citenamefont
  {{Deveaud-Pl{\'e}dran}}}]{manni_polariton_2011}%
  \BibitemOpen
  \bibfield  {author} {\bibinfo {author} {\bibfnamefont {F.}~\bibnamefont
  {Manni}}, \bibinfo {author} {\bibfnamefont {K.~G.}\ \bibnamefont
  {Lagoudakis}}, \bibinfo {author} {\bibfnamefont {B.}~\bibnamefont {Pietka}},
  \bibinfo {author} {\bibfnamefont {L.}~\bibnamefont {Fontanesi}}, \bibinfo
  {author} {\bibfnamefont {M.}~\bibnamefont {Wouters}}, \bibinfo {author}
  {\bibfnamefont {V.}~\bibnamefont {Savona}}, \bibinfo {author} {\bibfnamefont
  {R.}~\bibnamefont {Andr{\'e}}},\ and\ \bibinfo {author} {\bibfnamefont
  {B.}~\bibnamefont {{Deveaud-Pl{\'e}dran}}},\ }\bibfield  {title} {\bibinfo
  {title} {Polariton {{Condensation}} in a {{One-Dimensional Disordered
  Potential}}},\ }\href {https://doi.org/10.1103/PhysRevLett.106.176401}
  {\bibfield  {journal} {\bibinfo  {journal} {Phys. Rev. Lett.}\ }\textbf
  {\bibinfo {volume} {106}},\ \bibinfo {pages} {176401} (\bibinfo {year}
  {2011})}\BibitemShut {NoStop}%
\bibitem [{\citenamefont {Thunert}\ \emph {et~al.}(2016)\citenamefont
  {Thunert}, \citenamefont {Janot}, \citenamefont {Franke}, \citenamefont
  {Sturm}, \citenamefont {Michalsky}, \citenamefont {Mart{\'i}n}, \citenamefont
  {Vi{\~n}a}, \citenamefont {Rosenow}, \citenamefont {Grundmann},\ and\
  \citenamefont {{Schmidt-Grund}}}]{thunert_cavity_2016}%
  \BibitemOpen
  \bibfield  {author} {\bibinfo {author} {\bibfnamefont {M.}~\bibnamefont
  {Thunert}}, \bibinfo {author} {\bibfnamefont {A.}~\bibnamefont {Janot}},
  \bibinfo {author} {\bibfnamefont {H.}~\bibnamefont {Franke}}, \bibinfo
  {author} {\bibfnamefont {C.}~\bibnamefont {Sturm}}, \bibinfo {author}
  {\bibfnamefont {T.}~\bibnamefont {Michalsky}}, \bibinfo {author}
  {\bibfnamefont {M.~D.}\ \bibnamefont {Mart{\'i}n}}, \bibinfo {author}
  {\bibfnamefont {L.}~\bibnamefont {Vi{\~n}a}}, \bibinfo {author}
  {\bibfnamefont {B.}~\bibnamefont {Rosenow}}, \bibinfo {author} {\bibfnamefont
  {M.}~\bibnamefont {Grundmann}},\ and\ \bibinfo {author} {\bibfnamefont
  {R.}~\bibnamefont {{Schmidt-Grund}}},\ }\bibfield  {title} {\bibinfo {title}
  {Cavity polariton condensate in a disordered environment},\ }\href
  {https://doi.org/10.1103/PhysRevB.93.064203} {\bibfield  {journal} {\bibinfo
  {journal} {Phys. Rev. B}\ }\textbf {\bibinfo {volume} {93}},\ \bibinfo
  {pages} {064203} (\bibinfo {year} {2016})}\BibitemShut {NoStop}%
\bibitem [{\citenamefont {Janot}\ \emph {et~al.}(2013)\citenamefont {Janot},
  \citenamefont {Hyart}, \citenamefont {Eastham},\ and\ \citenamefont
  {Rosenow}}]{janot_superfluid_2013}%
  \BibitemOpen
  \bibfield  {author} {\bibinfo {author} {\bibfnamefont {A.}~\bibnamefont
  {Janot}}, \bibinfo {author} {\bibfnamefont {T.}~\bibnamefont {Hyart}},
  \bibinfo {author} {\bibfnamefont {P.~R.}\ \bibnamefont {Eastham}},\ and\
  \bibinfo {author} {\bibfnamefont {B.}~\bibnamefont {Rosenow}},\ }\bibfield
  {title} {\bibinfo {title} {Superfluid stiffness of a driven dissipative
  condensate with disorder},\ }\href
  {https://doi.org/10.1103/PhysRevLett.111.230403} {\bibfield  {journal}
  {\bibinfo  {journal} {Phys. Rev. Lett.}\ }\textbf {\bibinfo {volume} {111}},\
  \bibinfo {pages} {230403} (\bibinfo {year} {2013})}\BibitemShut {NoStop}%
\bibitem [{\citenamefont {Pikovskij}\ \emph {et~al.}(2003)\citenamefont
  {Pikovskij}, \citenamefont {Rosenblum},\ and\ \citenamefont
  {Kurths}}]{pikovskij_synchronization_2003}%
  \BibitemOpen
  \bibfield  {author} {\bibinfo {author} {\bibfnamefont {A.}~\bibnamefont
  {Pikovskij}}, \bibinfo {author} {\bibfnamefont {M.}~\bibnamefont
  {Rosenblum}},\ and\ \bibinfo {author} {\bibfnamefont {J.}~\bibnamefont
  {Kurths}},\ }\href@noop {} {\emph {\bibinfo {title} {Synchronization: A
  Universal Concept in Nonlinear Sciences}}},\ \bibinfo {series} {Cambridge
  Nonlinear Science Series}\ No.~\bibinfo {number} {12}\ (\bibinfo  {publisher}
  {{Cambridge University Press}},\ \bibinfo {address} {{Cambridge}},\ \bibinfo
  {year} {2003})\BibitemShut {NoStop}%
\bibitem [{\citenamefont {{Halpin-Healy}}\ and\ \citenamefont
  {Zhang}(1995)}]{halpin-healy_kinetic_1995}%
  \BibitemOpen
  \bibfield  {author} {\bibinfo {author} {\bibfnamefont {T.}~\bibnamefont
  {{Halpin-Healy}}}\ and\ \bibinfo {author} {\bibfnamefont {Y.-C.}\
  \bibnamefont {Zhang}},\ }\bibfield  {title} {\bibinfo {title} {Kinetic
  roughening phenomena, stochastic growth, directed polymers and all that.
  {{Aspects}} of multidisciplinary statistical mechanics},\ }\href
  {https://doi.org/10.1016/0370-1573(94)00087-J} {\bibfield  {journal}
  {\bibinfo  {journal} {Phy. Rep.}\ }\textbf {\bibinfo {volume} {254}},\
  \bibinfo {pages} {215} (\bibinfo {year} {1995})}\BibitemShut {NoStop}%
\bibitem [{\citenamefont {Manneville}\ and\ \citenamefont
  {Chat{\'e}}(1996)}]{manneville_phase_1996}%
  \BibitemOpen
  \bibfield  {author} {\bibinfo {author} {\bibfnamefont {P.}~\bibnamefont
  {Manneville}}\ and\ \bibinfo {author} {\bibfnamefont {H.}~\bibnamefont
  {Chat{\'e}}},\ }\bibfield  {title} {\bibinfo {title} {Phase turbulence in the
  two-dimensional complex {{Ginzburg-Landau}} equation},\ }\href
  {https://doi.org/10.1016/0167-2789(96)00045-0} {\bibfield  {journal}
  {\bibinfo  {journal} {Physica D}\ }\textbf {\bibinfo {volume} {96}},\
  \bibinfo {pages} {30} (\bibinfo {year} {1996})}\BibitemShut {NoStop}%
\bibitem [{\citenamefont {Kuramoto}(1984)}]{kuramoto_chemical_1984}%
  \BibitemOpen
  \bibfield  {author} {\bibinfo {author} {\bibfnamefont {Y.}~\bibnamefont
  {Kuramoto}},\ }\href@noop {} {\emph {\bibinfo {title} {Chemical
  {{Oscillations}}, {{Waves}}, and {{Turbulence}}}}}\ (\bibinfo  {publisher}
  {{Courier Corporation}},\ \bibinfo {year} {1984})\BibitemShut {NoStop}%
\bibitem [{\citenamefont {Kardar}\ \emph {et~al.}(1986)\citenamefont {Kardar},
  \citenamefont {Parisi},\ and\ \citenamefont {Zhang}}]{kardar_dynamic_1986}%
  \BibitemOpen
  \bibfield  {author} {\bibinfo {author} {\bibfnamefont {M.}~\bibnamefont
  {Kardar}}, \bibinfo {author} {\bibfnamefont {G.}~\bibnamefont {Parisi}},\
  and\ \bibinfo {author} {\bibfnamefont {Y.-C.}\ \bibnamefont {Zhang}},\
  }\bibfield  {title} {\bibinfo {title} {Dynamic scaling of growing
  interfaces},\ }\href {https://doi.org/10.1103/PhysRevLett.56.889} {\bibfield
  {journal} {\bibinfo  {journal} {Phys. Rev. Lett.}\ }\textbf {\bibinfo
  {volume} {56}},\ \bibinfo {pages} {889} (\bibinfo {year} {1986})}\BibitemShut
  {NoStop}%
\bibitem [{\citenamefont {Szendro}\ \emph {et~al.}(2007)\citenamefont
  {Szendro}, \citenamefont {L{\'o}pez},\ and\ \citenamefont
  {Rodr{\'i}guez}}]{szendro_localization_2007}%
  \BibitemOpen
  \bibfield  {author} {\bibinfo {author} {\bibfnamefont {I.~G.}\ \bibnamefont
  {Szendro}}, \bibinfo {author} {\bibfnamefont {J.~M.}\ \bibnamefont
  {L{\'o}pez}},\ and\ \bibinfo {author} {\bibfnamefont {M.~A.}\ \bibnamefont
  {Rodr{\'i}guez}},\ }\bibfield  {title} {\bibinfo {title} {Localization in
  disordered media, anomalous roughening, and coarsening dynamics of faceted
  surfaces},\ }\href {https://doi.org/10.1103/PhysRevE.76.011603} {\bibfield
  {journal} {\bibinfo  {journal} {Phys. Rev. E}\ }\textbf {\bibinfo {volume}
  {76}},\ \bibinfo {pages} {011603} (\bibinfo {year} {2007})}\BibitemShut
  {NoStop}%
\bibitem [{\citenamefont {Krug}\ and\ \citenamefont
  {{Halpin-Healy}}(1993)}]{krug_directed_1993}%
  \BibitemOpen
  \bibfield  {author} {\bibinfo {author} {\bibfnamefont {J.}~\bibnamefont
  {Krug}}\ and\ \bibinfo {author} {\bibfnamefont {T.}~\bibnamefont
  {{Halpin-Healy}}},\ }\bibfield  {title} {\bibinfo {title} {Directed polymers
  in the presence of columnar disorder},\ }\href
  {https://doi.org/10.1051/jp1:1993240} {\bibfield  {journal} {\bibinfo
  {journal} {J. Phys. I France}\ }\textbf {\bibinfo {volume} {3}},\ \bibinfo
  {pages} {2179} (\bibinfo {year} {1993})}\BibitemShut {NoStop}%
\bibitem [{\citenamefont {Nattermann}\ and\ \citenamefont
  {Renz}(1989)}]{nattermann_diffusion_1989}%
  \BibitemOpen
  \bibfield  {author} {\bibinfo {author} {\bibfnamefont {T.}~\bibnamefont
  {Nattermann}}\ and\ \bibinfo {author} {\bibfnamefont {W.}~\bibnamefont
  {Renz}},\ }\bibfield  {title} {\bibinfo {title} {Diffusion in a random
  catalytic environment, polymers in random media, and stochastically growing
  interfaces},\ }\href {https://doi.org/10.1103/PhysRevA.40.4675} {\bibfield
  {journal} {\bibinfo  {journal} {Phys. Rev. A}\ }\textbf {\bibinfo {volume}
  {40}},\ \bibinfo {pages} {4675} (\bibinfo {year} {1989})}\BibitemShut
  {NoStop}%
\bibitem [{\citenamefont {Blasius}\ and\ \citenamefont
  {T{\"o}njes}(2005)}]{blasius_quasiregular_2005}%
  \BibitemOpen
  \bibfield  {author} {\bibinfo {author} {\bibfnamefont {B.}~\bibnamefont
  {Blasius}}\ and\ \bibinfo {author} {\bibfnamefont {R.}~\bibnamefont
  {T{\"o}njes}},\ }\bibfield  {title} {\bibinfo {title} {Quasiregular
  {{Concentric Waves}} in {{Heterogeneous Lattices}} of {{Coupled
  Oscillators}}},\ }\href {https://doi.org/10.1103/PhysRevLett.95.084101}
  {\bibfield  {journal} {\bibinfo  {journal} {Phys. Rev. Lett.}\ }\textbf
  {\bibinfo {volume} {95}},\ \bibinfo {pages} {084101} (\bibinfo {year}
  {2005})}\BibitemShut {NoStop}%
\bibitem [{\citenamefont {Sakaguchi}\ \emph {et~al.}(1988)\citenamefont
  {Sakaguchi}, \citenamefont {Shinomoto},\ and\ \citenamefont
  {Kuramoto}}]{sakaguchi_mutual_1988}%
  \BibitemOpen
  \bibfield  {author} {\bibinfo {author} {\bibfnamefont {H.}~\bibnamefont
  {Sakaguchi}}, \bibinfo {author} {\bibfnamefont {S.}~\bibnamefont
  {Shinomoto}},\ and\ \bibinfo {author} {\bibfnamefont {Y.}~\bibnamefont
  {Kuramoto}},\ }\bibfield  {title} {\bibinfo {title} {Mutual {{Entrainment}}
  in {{Oscillator Lattices}} with {{Nonvariational Type Interaction}}},\ }\href
  {https://doi.org/10.1143/PTP.79.1069} {\bibfield  {journal} {\bibinfo
  {journal} {Prog. Theor. Phys.}\ }\textbf {\bibinfo {volume} {79}},\ \bibinfo
  {pages} {1069} (\bibinfo {year} {1988})}\BibitemShut {NoStop}%
\bibitem [{\citenamefont {Guti{\'e}rrez}\ and\ \citenamefont
  {Cuerno}(2023)}]{gutierrez_nonequilibrium_2023}%
  \BibitemOpen
  \bibfield  {author} {\bibinfo {author} {\bibfnamefont {R.}~\bibnamefont
  {Guti{\'e}rrez}}\ and\ \bibinfo {author} {\bibfnamefont {R.}~\bibnamefont
  {Cuerno}},\ }\bibfield  {title} {\bibinfo {title} {Nonequilibrium criticality
  driven by {{Kardar-Parisi-Zhang}} fluctuations in the synchronization of
  oscillator lattices},\ }\href
  {https://doi.org/10.1103/PhysRevResearch.5.023047} {\bibfield  {journal}
  {\bibinfo  {journal} {Phys. Rev. Res.}\ }\textbf {\bibinfo {volume} {5}},\
  \bibinfo {pages} {023047} (\bibinfo {year} {2023})}\BibitemShut {NoStop}%
\bibitem [{\citenamefont {Lauter}\ \emph {et~al.}(2017)\citenamefont {Lauter},
  \citenamefont {Mitra},\ and\ \citenamefont
  {Marquardt}}]{lauter_kardar-parisi-zhang_2017}%
  \BibitemOpen
  \bibfield  {author} {\bibinfo {author} {\bibfnamefont {R.}~\bibnamefont
  {Lauter}}, \bibinfo {author} {\bibfnamefont {A.}~\bibnamefont {Mitra}},\ and\
  \bibinfo {author} {\bibfnamefont {F.}~\bibnamefont {Marquardt}},\ }\bibfield
  {title} {\bibinfo {title} {From {{Kardar-Parisi-Zhang}} scaling to explosive
  desynchronization in arrays of limit-cycle oscillators},\ }\href
  {https://doi.org/10.1103/PhysRevE.96.012220} {\bibfield  {journal} {\bibinfo
  {journal} {Phys. Rev. E}\ }\textbf {\bibinfo {volume} {96}},\ \bibinfo
  {pages} {012220} (\bibinfo {year} {2017})}\BibitemShut {NoStop}%
\bibitem [{\citenamefont {Lauter}\ \emph {et~al.}(2015)\citenamefont {Lauter},
  \citenamefont {Brendel}, \citenamefont {Habraken},\ and\ \citenamefont
  {Marquardt}}]{lauter_pattern_2015}%
  \BibitemOpen
  \bibfield  {author} {\bibinfo {author} {\bibfnamefont {R.}~\bibnamefont
  {Lauter}}, \bibinfo {author} {\bibfnamefont {C.}~\bibnamefont {Brendel}},
  \bibinfo {author} {\bibfnamefont {S.~J.~M.}\ \bibnamefont {Habraken}},\ and\
  \bibinfo {author} {\bibfnamefont {F.}~\bibnamefont {Marquardt}},\ }\bibfield
  {title} {\bibinfo {title} {Pattern phase diagram for two-dimensional arrays
  of coupled limit-cycle oscillators},\ }\href
  {https://doi.org/10.1103/PhysRevE.92.012902} {\bibfield  {journal} {\bibinfo
  {journal} {Phys. Rev. E}\ }\textbf {\bibinfo {volume} {92}},\ \bibinfo
  {pages} {012902} (\bibinfo {year} {2015})}\BibitemShut {NoStop}%
\bibitem [{\citenamefont {Sieberer}\ and\ \citenamefont
  {Altman}(2018)}]{sieberer_topological_2018}%
  \BibitemOpen
  \bibfield  {author} {\bibinfo {author} {\bibfnamefont {L.~M.}\ \bibnamefont
  {Sieberer}}\ and\ \bibinfo {author} {\bibfnamefont {E.}~\bibnamefont
  {Altman}},\ }\bibfield  {title} {\bibinfo {title} {Topological {{Defects}} in
  {{Anisotropic Driven Open Systems}}},\ }\href
  {https://doi.org/10.1103/PhysRevLett.121.085704} {\bibfield  {journal}
  {\bibinfo  {journal} {Phys. Rev. Lett.}\ }\textbf {\bibinfo {volume} {121}},\
  \bibinfo {pages} {085704} (\bibinfo {year} {2018})}\BibitemShut {NoStop}%
\bibitem [{\citenamefont {Caputo}\ \emph {et~al.}(2018)\citenamefont {Caputo},
  \citenamefont {Ballarini}, \citenamefont {Dagvadorj}, \citenamefont
  {S{\'a}nchez~Mu{\~n}oz}, \citenamefont {De~Giorgi}, \citenamefont {Dominici},
  \citenamefont {West}, \citenamefont {Pfeiffer}, \citenamefont {Gigli},
  \citenamefont {Laussy}, \citenamefont {Szyma{\'n}ska},\ and\ \citenamefont
  {Sanvitto}}]{caputo_topological_2018}%
  \BibitemOpen
  \bibfield  {author} {\bibinfo {author} {\bibfnamefont {D.}~\bibnamefont
  {Caputo}}, \bibinfo {author} {\bibfnamefont {D.}~\bibnamefont {Ballarini}},
  \bibinfo {author} {\bibfnamefont {G.}~\bibnamefont {Dagvadorj}}, \bibinfo
  {author} {\bibfnamefont {C.}~\bibnamefont {S{\'a}nchez~Mu{\~n}oz}}, \bibinfo
  {author} {\bibfnamefont {M.}~\bibnamefont {De~Giorgi}}, \bibinfo {author}
  {\bibfnamefont {L.}~\bibnamefont {Dominici}}, \bibinfo {author}
  {\bibfnamefont {K.}~\bibnamefont {West}}, \bibinfo {author} {\bibfnamefont
  {L.~N.}\ \bibnamefont {Pfeiffer}}, \bibinfo {author} {\bibfnamefont
  {G.}~\bibnamefont {Gigli}}, \bibinfo {author} {\bibfnamefont {F.~P.}\
  \bibnamefont {Laussy}}, \bibinfo {author} {\bibfnamefont {M.~H.}\
  \bibnamefont {Szyma{\'n}ska}},\ and\ \bibinfo {author} {\bibfnamefont
  {D.}~\bibnamefont {Sanvitto}},\ }\bibfield  {title} {\bibinfo {title}
  {Topological order and thermal equilibrium in polariton condensates},\ }\href
  {https://doi.org/10.1038/nmat5039} {\bibfield  {journal} {\bibinfo  {journal}
  {Nat. Mater.}\ }\textbf {\bibinfo {volume} {17}},\ \bibinfo {pages} {145}
  (\bibinfo {year} {2018})}\BibitemShut {NoStop}%
\bibitem [{\citenamefont {He}\ \emph {et~al.}(2017)\citenamefont {He},
  \citenamefont {Sieberer},\ and\ \citenamefont {Diehl}}]{he_space-time_2017}%
  \BibitemOpen
  \bibfield  {author} {\bibinfo {author} {\bibfnamefont {L.}~\bibnamefont
  {He}}, \bibinfo {author} {\bibfnamefont {L.~M.}\ \bibnamefont {Sieberer}},\
  and\ \bibinfo {author} {\bibfnamefont {S.}~\bibnamefont {Diehl}},\ }\bibfield
   {title} {\bibinfo {title} {Space-time vortex driven crossover and vortex
  turbulence phase transition in one-dimensional driven open condensates},\
  }\href {https://doi.org/10.1103/PhysRevLett.118.085301} {\bibfield  {journal}
  {\bibinfo  {journal} {Phys. Rev. Lett.}\ }\textbf {\bibinfo {volume} {118}},\
  \bibinfo {pages} {085301} (\bibinfo {year} {2017})}\BibitemShut {NoStop}%
\bibitem [{\citenamefont {Ohadi}\ \emph {et~al.}(2018)\citenamefont {Ohadi},
  \citenamefont {{del Valle-Inclan Redondo}}, \citenamefont {Ramsay},
  \citenamefont {Hatzopoulos}, \citenamefont {Liew}, \citenamefont {Eastham},
  \citenamefont {Savvidis},\ and\ \citenamefont
  {Baumberg}}]{ohadi_synchronization_2018}%
  \BibitemOpen
  \bibfield  {author} {\bibinfo {author} {\bibfnamefont {H.}~\bibnamefont
  {Ohadi}}, \bibinfo {author} {\bibfnamefont {Y.}~\bibnamefont {{del
  Valle-Inclan Redondo}}}, \bibinfo {author} {\bibfnamefont {A.~J.}\
  \bibnamefont {Ramsay}}, \bibinfo {author} {\bibfnamefont {Z.}~\bibnamefont
  {Hatzopoulos}}, \bibinfo {author} {\bibfnamefont {T.~C.~H.}\ \bibnamefont
  {Liew}}, \bibinfo {author} {\bibfnamefont {P.~R.}\ \bibnamefont {Eastham}},
  \bibinfo {author} {\bibfnamefont {P.~G.}\ \bibnamefont {Savvidis}},\ and\
  \bibinfo {author} {\bibfnamefont {J.~J.}\ \bibnamefont {Baumberg}},\
  }\bibfield  {title} {\bibinfo {title} {Synchronization crossover of polariton
  condensates in weakly disordered lattices},\ }\href
  {https://doi.org/10.1103/PhysRevB.97.195109} {\bibfield  {journal} {\bibinfo
  {journal} {Phys. Rev. B}\ }\textbf {\bibinfo {volume} {97}},\ \bibinfo
  {pages} {195109} (\bibinfo {year} {2018})}\BibitemShut {NoStop}%
\bibitem [{\citenamefont {Aleiner}\ \emph {et~al.}(2012)\citenamefont
  {Aleiner}, \citenamefont {Altshuler},\ and\ \citenamefont
  {Rubo}}]{aleiner_radiative_2012}%
  \BibitemOpen
  \bibfield  {author} {\bibinfo {author} {\bibfnamefont {I.~L.}\ \bibnamefont
  {Aleiner}}, \bibinfo {author} {\bibfnamefont {B.~L.}\ \bibnamefont
  {Altshuler}},\ and\ \bibinfo {author} {\bibfnamefont {Y.~G.}\ \bibnamefont
  {Rubo}},\ }\bibfield  {title} {\bibinfo {title} {Radiative coupling and weak
  lasing of exciton-polariton condensates},\ }\href
  {https://doi.org/10.1103/PhysRevB.85.121301} {\bibfield  {journal} {\bibinfo
  {journal} {Phys. Rev. B}\ }\textbf {\bibinfo {volume} {85}},\ \bibinfo
  {pages} {121301(R)} (\bibinfo {year} {2012})}\BibitemShut {NoStop}%
\bibitem [{\citenamefont {T{\"o}pfer}\ \emph {et~al.}(2021)\citenamefont
  {T{\"o}pfer}, \citenamefont {T{\"o}pfer}, \citenamefont {Chatzopoulos},
  \citenamefont {Sigurdsson}, \citenamefont {Sigurdsson}, \citenamefont
  {Cookson}, \citenamefont {Rubo}, \citenamefont {Lagoudakis},\ and\
  \citenamefont {Lagoudakis}}]{topfer_engineering_2021}%
  \BibitemOpen
  \bibfield  {author} {\bibinfo {author} {\bibfnamefont {J.~D.}\ \bibnamefont
  {T{\"o}pfer}}, \bibinfo {author} {\bibfnamefont {J.~D.}\ \bibnamefont
  {T{\"o}pfer}}, \bibinfo {author} {\bibfnamefont {I.}~\bibnamefont
  {Chatzopoulos}}, \bibinfo {author} {\bibfnamefont {H.}~\bibnamefont
  {Sigurdsson}}, \bibinfo {author} {\bibfnamefont {H.}~\bibnamefont
  {Sigurdsson}}, \bibinfo {author} {\bibfnamefont {T.}~\bibnamefont {Cookson}},
  \bibinfo {author} {\bibfnamefont {Y.~G.}\ \bibnamefont {Rubo}}, \bibinfo
  {author} {\bibfnamefont {P.~G.}\ \bibnamefont {Lagoudakis}},\ and\ \bibinfo
  {author} {\bibfnamefont {P.~G.}\ \bibnamefont {Lagoudakis}},\ }\bibfield
  {title} {\bibinfo {title} {Engineering spatial coherence in lattices of
  polariton condensates},\ }\href {https://doi.org/10.1364/OPTICA.409976}
  {\bibfield  {journal} {\bibinfo  {journal} {Optica}\ }\textbf {\bibinfo
  {volume} {8}},\ \bibinfo {pages} {106} (\bibinfo {year} {2021})}\BibitemShut
  {NoStop}%
\bibitem [{\citenamefont {Wertz}\ \emph {et~al.}(2010)\citenamefont {Wertz},
  \citenamefont {Ferrier}, \citenamefont {Solnyshkov}, \citenamefont {Johne},
  \citenamefont {Sanvitto}, \citenamefont {Lema{\^i}tre}, \citenamefont
  {Sagnes}, \citenamefont {Grousson}, \citenamefont {Kavokin}, \citenamefont
  {Senellart}, \citenamefont {Malpuech},\ and\ \citenamefont
  {Bloch}}]{wertz_spontaneous_2010}%
  \BibitemOpen
  \bibfield  {author} {\bibinfo {author} {\bibfnamefont {E.}~\bibnamefont
  {Wertz}}, \bibinfo {author} {\bibfnamefont {L.}~\bibnamefont {Ferrier}},
  \bibinfo {author} {\bibfnamefont {D.~D.}\ \bibnamefont {Solnyshkov}},
  \bibinfo {author} {\bibfnamefont {R.}~\bibnamefont {Johne}}, \bibinfo
  {author} {\bibfnamefont {D.}~\bibnamefont {Sanvitto}}, \bibinfo {author}
  {\bibfnamefont {A.}~\bibnamefont {Lema{\^i}tre}}, \bibinfo {author}
  {\bibfnamefont {I.}~\bibnamefont {Sagnes}}, \bibinfo {author} {\bibfnamefont
  {R.}~\bibnamefont {Grousson}}, \bibinfo {author} {\bibfnamefont {A.~V.}\
  \bibnamefont {Kavokin}}, \bibinfo {author} {\bibfnamefont {P.}~\bibnamefont
  {Senellart}}, \bibinfo {author} {\bibfnamefont {G.}~\bibnamefont
  {Malpuech}},\ and\ \bibinfo {author} {\bibfnamefont {J.}~\bibnamefont
  {Bloch}},\ }\bibfield  {title} {\bibinfo {title} {Spontaneous formation and
  optical manipulation of extended polariton condensates},\ }\href
  {https://doi.org/10.1038/nphys1750} {\bibfield  {journal} {\bibinfo
  {journal} {Nat. Phys.}\ }\textbf {\bibinfo {volume} {6}},\ \bibinfo {pages}
  {860} (\bibinfo {year} {2010})}\BibitemShut {NoStop}%
\bibitem [{\citenamefont {T{\"o}pfer}\ \emph {et~al.}(2020)\citenamefont
  {T{\"o}pfer}, \citenamefont {Sigurdsson}, \citenamefont {Pickup},\ and\
  \citenamefont {Lagoudakis}}]{topfer_time-delay_2020}%
  \BibitemOpen
  \bibfield  {author} {\bibinfo {author} {\bibfnamefont {J.~D.}\ \bibnamefont
  {T{\"o}pfer}}, \bibinfo {author} {\bibfnamefont {H.}~\bibnamefont
  {Sigurdsson}}, \bibinfo {author} {\bibfnamefont {L.}~\bibnamefont {Pickup}},\
  and\ \bibinfo {author} {\bibfnamefont {P.~G.}\ \bibnamefont {Lagoudakis}},\
  }\bibfield  {title} {\bibinfo {title} {Time-delay polaritonics},\ }\href
  {https://doi.org/10.1038/s42005-019-0271-0} {\bibfield  {journal} {\bibinfo
  {journal} {Commun. Phys.}\ }\textbf {\bibinfo {volume} {3}},\ \bibinfo
  {pages} {2} (\bibinfo {year} {2020})}\BibitemShut {NoStop}%
\bibitem [{\citenamefont {R{\"o}{\ss}ler}(2004)}]{rosler_rungekutta_2004}%
  \BibitemOpen
  \bibfield  {author} {\bibinfo {author} {\bibfnamefont {A.}~\bibnamefont
  {R{\"o}{\ss}ler}},\ }\bibfield  {title} {\bibinfo {title}
  {Runge\textendash{{Kutta}} methods for {{Stratonovich}} stochastic
  differential equation systems with commutative noise},\ }\href
  {https://doi.org/10.1016/j.cam.2003.09.009} {\bibfield  {journal} {\bibinfo
  {journal} {J. Comput. Appl. Math.}\ }\textbf {\bibinfo {volume} {164--165}},\
  \bibinfo {pages} {613} (\bibinfo {year} {2004})}\BibitemShut {NoStop}%
\bibitem [{\citenamefont {Stratonovich}(1967)}]{stratonovich_topics_1967}%
  \BibitemOpen
  \bibfield  {author} {\bibinfo {author} {\bibfnamefont {R.~L.}\ \bibnamefont
  {Stratonovich}},\ }\href@noop {} {\emph {\bibinfo {title} {Topics in the
  {{Theory}} of {{Random Noise}}}}},\ Vol.~\bibinfo {volume} {2}\ (\bibinfo
  {publisher} {{Gordon and Breach}},\ \bibinfo {address} {{New York}},\
  \bibinfo {year} {1967})\BibitemShut {NoStop}%
\end{thebibliography}%

\end{document}